\documentclass[useAMS,usenatbib]{mn2e}
\usepackage{graphicx}
\usepackage{amsmath}
\usepackage{amssymb,latexsym,amsopn}
\title[Gravitational waves from binary SMBHs]{Binary super-massive black hole environments 
diminish the gravitational-wave signal in the pulsar timing band}

\author[V. Ravi et al.]{V. Ravi,$^{1, 2}$\thanks{E-mail: v.vikram.ravi@gmail.com} 
J. S. B. Wyithe,$^{1}$
R. M. Shannon,$^{2}$ 
G. Hobbs$^{2}$
and R. N. Manchester$^{2}$
\\
$^{1}$School of Physics, University of Melbourne, Parkville VIC 3010, Australia\\
$^{2}$CSIRO Astronomy and Space Science, Australia Telescope National Facility, P.O. Box 76, Epping, NSW 1710, Australia\\
}

\begin{document}

\date{}

\pagerange{1--15} \pubyear{2014}

\maketitle

\label{firstpage}

\begin{abstract}

We assess the effects of super-massive black hole (SMBH) environments on the gravitational-wave (GW) signal 
from binary SMBHs. To date, searches with pulsar timing arrays for GWs from binary SMBHs, in the 
frequency band $\sim1-100$\,nHz, include the assumptions that all binaries are circular and evolve 
only through GW emission. However, dynamical studies have shown that the only way that binary SMBH orbits can decay to 
separations where GW emission dominates the evolution is through interactions with their environments. 
We augment an existing galaxy and SMBH formation and evolution model with calculations of binary SMBH evolution in stellar 
environments, accounting for non-zero binary eccentricities. We find that coupling between binaries and their environments 
causes the expected GW spectral energy distribution to be reduced with respect to the standard assumption of circular, 
GW-driven binaries, for frequencies up to $\sim20$\,nHz. Larger eccentricities at binary formation further reduce the signal 
in this regime. We also find that GW bursts from individual eccentric binary SMBHs are unlikely to be detectable with current 
pulsar timing arrays. The uncertainties in these predictions are large, owing to observational uncertainty in SMBH-galaxy scaling 
relations and the galaxy stellar mass function, uncertainty in the nature of binary-environment coupling, and uncertainty in the 
numbers of the most massive binary SMBHs. We conclude, however, that low-frequency GWs from binary SMBHs may be more 
difficult to detect with pulsar timing arrays than currently thought.

\end{abstract}


\section{Introduction} 

The merger of a pair of galaxies hosting central super-massive black holes (SMBHs) is expected to result in the formation of a binary SMBH 
\nocite{bbr80}(Begelman, Blandford \& Rees 1980). The central SMBHs sink in the merger remnant potential well through the action of dynamical friction, 
and form a bound binary when the mass within the orbit of the lighter SMBH is dominated by the heavier SMBH. As stars within 
the binary orbit are quickly ejected, the binary will decay further only 
if another mechanism to extract binding energy and angular momentum exists. Proposed mechanisms include slingshot scattering of 
stars on radial, low angular momentum orbits intersecting the binary \citep{fr76,q96,y02}, and friction against a spherical Bondi gas accretion flow 
\citep{elc+04} or a circum-nuclear gas disk \citep[e.g.,][]{rds+11}. If the orbital decay process can drive the binary to a small separation, 
gravitational-wave (GW) emission will eventually cause the binary to coalesce \citep[e.g.,][]{pm63,bcc+06}.   

In a cosmological context, merging dark matter halos follow parabolic trajectories \citep[e.g.,][]{vll+99,w11}, implying 
large initial eccentricities \citep[typically $\sim0.6$,][]{hfm03} for the orbits of SMBHs sinking towards galaxy merger remnant centres. Steep stellar 
density gradients in merging galaxies may reduce this eccentricity; indeed, some models suggest that binary SMBHs are 
likely to be close to circular upon formation \nocite{cpv87,hfm03}\citep[Casertano, Phinney \& Villumsen 1987; Hashimoto, Funato \& Makino 2007;][]{pr94}. Slingshot interactions between binaries and individual stars again grow 
the eccentricities \nocite{shm06}\citep[e.g., Sesana, Haardt \& Madau 2006;][]{q96,bpb+09,kpb+12}, because binaries spend more time, and hence lose more energy, at larger 
separations. \citet{rds+11} found that binary SMBHs embedded in massive self-gravitating gas disks will have large eccentricities, 
between 0.6 and 0.8, at the onset of GW-dominated evolution.  In the GW-dominated regime, however, binaries are expected to quickly circularise 
\citep[e.g.,][]{pm63,bcc+06}. 

There is no direct observational evidence for the existence of binary SMBHs 
\nocite{dsd12}(Dotti, Sesana \& Decarli 2012). However, the GW emission from binaries prior to coalescence is an unambiguous signature of their 
existence. Observing GWs from binary SMBHs will enable binary SMBH physics, as well as models for the formation and evolution of the 
cosmological SMBH population, to be observationally 
tested. Here, we focus on the possibility of detecting GWs from binary SMBHs in the early parts of their GW-dominated 
evolutionary stages with radio pulsar timing arrays \citep[PTAs;][]{fb90,mhb+13}. PTAs are currently sensitive to GWs in the 
frequency band $\sim1-100$\,nHz, which is complementary to other GW detection experiments.

PTAs target both a stochastic, isotropic background of GWs from binary SMBHs \citep[e.g.,][]{ych+11,vlj+11} and GWs from individual binary 
systems \nocite{esc12,yhj+10}(Yardley et al. 2010; Ellis, Siemens \& Creighton 2012). The summed GW signal from all binary SMBHs in the Universe is 
expected to approximate an isotropic background, although individual binaries are potentially detectable at all frequencies within the PTA 
band \citep{rwh+12}. Recent PTA results suggest that a large fraction of existing models for the GW background from binary 
SMBHs is inconsistent with observations \citep{src+13}. However, most current predictions 
for the spectral shape \citep{p01}, statistical nature \nocite{svv09}\citep[Sesana, Vecchio \& Volonteri 2009;][]{rwh+12} and strength \citep{s12} of the GW signal from binary 
SMBHs assume that all binaries are in circular orbits, and losing energy and angular momentum only to GWs. These assumptions 
correspond to the well-known power law GW background characteristic strain spectrum from binary SMBHs 
that is proportional to $f^{-2/3}$, where $f$ is the GW frequency.

Here, we present an examination of the properties of the GW signal from binary SMBHs given a realistic model for binary orbital 
evolution. We use a semi-analytic model for galaxy and SMBH formation and evolution \citep[][hereafter G11]{gwb+11} implemented in the 
Millennium simulation \citep{swj+05} to specify the 
coalescence rate of binary SMBHs, and augment this with a framework \citep{s10} for the evolution of binary SMBHs in stellar environments. 
We neglect gas-driven binary evolution. This is because massive galaxies at low redshifts, which are expected to dominate the GW signal from binary 
SMBHs, will typically be late-type and gas-poor \nocite{ylm+11,mop12}(e.g., Yu et al. 2011; McWilliams, Ostriker \& Pretorius 2012).

Two key phenomena in binary SMBH evolution affect the summed GW signal relative to the case of circular binaries evolving under 
GW emission alone\footnote{We refer to this as the ``circular, GW-driven case'' throughout the paper.}:
\begin{enumerate}
\item Interactions between binary SMBHs and their environments will accelerate orbital decay compared to purely GW-driven binaries, 
reducing the time each binary spends radiating GWs. This may reduce the energy density in GWs at the lower end of the 
PTA frequency band \citep[e.g.,][]{wl03,s13}. 
\item While circular binaries emit GWs at the second harmonics of their orbital frequencies, eccentric binaries emit GWs at multiple 
harmonics \citep{pm63}. Given a population of binary SMBHs, this is expected to transfer GW energy density from lower 
frequencies in the PTA frequency band to higher frequencies \citep{en07,s13}. 
\end{enumerate}
We consider the effects of both these phenomena on the GW signal from binary SMBHs relative to the circular, GW-driven case. 
We also examine the possibility of detecting bursts of GWs from individual eccentric, massive binaries. In \S2, we outline the binary 
population model. We give our predictions for the summed GW signal in \S3, along with a discussion of uncertainties in our model. 
We consider the possibility of detectable GW bursts in \S4. Finally, we summarise our results in \S5 and present our conclusions \S6. 
Summaries of key PTA implications can be found at the ends of \S3, \S4 and \S5. We adopt a concordance cosmology 
consistent with the Millennium simulation \citep{swj+05}, with $\Omega_{M}=0.25$, $\Omega_{b}=0.045$, $\Omega_{\Lambda}=0.75$, and 
$H_{0}= 73$\,km\,s$^{-1}$\,Mpc$^{-1}$.

\section{Description of modelling methods}

\subsection{The gravitational wave background from a cosmological source population}

Consider a population of GW sources with comoving volume density $N(z)$ at redshift $z$, each radiating a GW luminosity per unit 
rest-frame frequency, $f_{r}$, of $L(f_{r})$. The specific intensity of GWs at 
the Earth from sources between redshifts $z$ and $z+dz$ is 
\begin{equation}
dI = \frac{L(f_{r})}{4\pi d_{L}^{2}}\frac{df_{r}}{df}N(z)\frac{d^{2}V_{c}}{d\Omega dz}dz.
\end{equation}
Here, $\frac{df_{r}}{df}=(1+z)$, where $f$ is the observed GW frequency, and the comoving volume element is 
\begin{displaymath}
\frac{d^{2}V_{c}}{d\Omega dz}=\frac{cd_{L}^{2}}{H(z)(1+z)^{2}},
\end{displaymath}
where the Hubble parameter, $H(z)$, is given by $H(z)=H_{0}[\Omega_{M}(1+z)^{3}+\Omega_{\Lambda}]^{1/2}$, $c$ is the vacuum 
speed of light, and $d_{L}$ is the luminosity distance at redshift $z$.
Now, following \citet{p01} and re-arranging, the energy density in GWs at the Earth per logarithmic frequency unit is, in any homogeneous 
and isotropic universe,
\begin{eqnarray}
\Omega_{\text{GW}}(f)\rho_{c}c^{2}&=&\frac{4\pi}{c}f\int dI \\
&=& \int_{0}^{\infty}\frac{fL(f_{r})N(z)}{H(z)(1+z)}dz \\
&=& \int_{t_{r}(\infty)}^{0}\frac{f_{r}}{1+z}L(f_{r})N(z)dt_{r}
\end{eqnarray}
where $f_{r}=f(1+z)$, $\rho_{c}=3H_{0}^{2}/(8\pi G)$ (with $G$ as the gravitational constant) is the critical mass density of the Universe, 
and $t_{r}$ is the proper time. The redshift $z$ is related to $t_{r}$ as 
\begin{equation}
t_{r}(z)=\int_{z}^{0}\frac{1}{H(z')(1+z')}dz'.
\end{equation} 
Although our Equation~(4) is directly comparable to Equation~(5) of \citet{p01}, we note that ours and Phinney's expressions are only mathematically 
identical if we explicitly assume that each source radiates for an infinitesimal (proper) time.

\subsection{The binary SMBH population at formation}

To begin, we consider binary SMBHs with component masses $M_{1}\geq M_{2}$, orbital semi-major axes $a_{0}$ and 
eccentricities $e_{0}$, embedded in isotropic, unbound cuspy stellar distributions with velocity dispersions $\sigma$. \citet{q96} 
found that binary hardening caused by slingshot interactions with individual stars becomes effective at binary component separations of 
\begin{equation}
a_{h}\approx \frac{GM_{2}}{4\sigma^{2}},
\end{equation} 
We assume that dynamical friction is effective in driving binaries to mean separations $a_{h}$ \citep[e.g.,][]{cmk+09,kpb+12}, and 
consider binaries at this stage to be newly formed.

We adopt a simple, one-parameter distribution for the eccentricities of binary SMBHs at formation, 
based on the postulate that the semi-major and semi-minor axes of the orbits ($a_{0}$ and $b_{0}$ respectively) are each log-normally distributed. 
This is justified because (\textit{a}) while many stellar encounters influence the values of $a_{0}$ and $b_{0}$, the effects of these encounters 
on the parameter values are heterogeneous, and (\textit{b}) both $a_{0}$ and $b_{0}$ are strictly positive.\footnote{See \citet{g45} for a discussion of the 
ubiquity of log-normal distributions in nature.} In general, log-normal distributions are used to model positive-definite random variables that are influenced by 
many multiplicative effects of differing magnitudes (i.e., heterogeneous effects). The central limit theorem implies that the \textit{product} of a large number of 
finite-variance positive random variables will approximately have a log-normal distribution.

We hence model the ratio $b_{0}/a_{0}$ using a probability density function given by 
\begin{equation}
F_{0}\left(\frac{b_{0}}{a_{0}},w_{0}\right) = \begin{cases}
\sqrt{\frac{2}{\pi}}\frac{a_{0}}{b_{0}w_{0}}\text{exp}\left[-\left(\frac{\ln(\frac{b_{0}}{a_{0}})}{w_{0}\sqrt{2}}\right)^{2}\right], & \text{$\frac{b_{0}}{a_{0}}\leq1$} \\
0, & \text{otherwise}
\end{cases} 
\end{equation}
Here, $w_{0}$ is the free parameter; larger values of $w_{0}$ correspond to typically larger binary eccentricities, and $w_{0}=0$ corresponds 
to a population of circular binaries. 
We do not consider any variation of $w_{0}$ with binary component masses or redshift, 
because there are no strong motivations for such variations. The eccentricity of a binary at $a_{h}$ is given by 
$e_{0}=\sqrt{1-(b_{0}/a_{0})^{2}}$.

Let $\mathbf{\zeta_{0}}=[M_{1},M_{2},e_{0}]$ be a vector of parameters of binaries at formation. 
We denote the distribution of binaries in these parameters as 
$D_{\mathbf{\zeta_{0}}}[N(\mathbf{\zeta_{0}},z)]$. In this notation, the multivariate density function for a parameter vector 
$\mathbf{X}$ with components $X_{i}$ indexed by an integer $i$ is given by 
\begin{displaymath}
D_{\mathbf{X}}[N]\equiv \prod_{i}\frac{\partial [N]}{\partial X_{i}}.
\end{displaymath}
Binaries at formation have semi-major axes $a_{0}=a_{h}/(1+\sqrt{1-e_{0}^{2}})$. 

We use the results of the semi-analytic model of G11 to specify $D_{\mathbf{\zeta_{0}}}[N(\mathbf{\zeta_{0}},z)]$. 
As outlined in \citet{rwh+12} and \citet{src+13}, the G11 results can be used to predict the coalescence rate of binary SMBHs. 
For this work, we only use coalescences with both $M_{1}$ and $M_{2}$ greater than $10^{6}M_{\odot}$, and 
only draw from the implementation of the G11 model in the Millennium simulation. We scale all SMBH masses by a factor of 
1.9 \citep{src+13} to account for recent SMBH and galaxy bulge measurements. 

We count the coalescing pairs of SMBHs in bins of $z$, $M_{1}$ and $M_{2}$ (with widths $\Delta z$, $\Delta M_{1}$, and $\Delta M_{2}$ respectively) 
within the entire Millennium simulation box. Binaries are also randomly assigned values of $e_{0}$ using Equation~(7). In this work, 
we consider four different initial binary eccentricity distributions defined by $w_{0}=0,\,0.1,\,0.35,\,0.93$.  
Denoting the binary counts for different values of $z$, $M_{1}$, $M_{2}$ and $e_{0}$ by the discrete distribution $n(\mathbf{\zeta_{0}},z)$, we 
have 
\begin{equation}
\frac{d}{dz}[D_{\mathbf{\zeta_{0}}}[N(\mathbf{\zeta_{0}},z)]]\approx\frac{n(z,\mathbf{\zeta_{0}})}{V_{\text{Mil}}\Delta z \Delta M_{1} \Delta M_{2} \Delta e_{0}},
\end{equation}
where $V_{\text{Mil}}$ is the comoving volume of the Millennium simulation box \citep{swj+05}. We average the distribution $n(\mathbf{\zeta_{0}},z)$ 
over merger order \citep{rwh+12} and 1000 realisations of the initial $e_{0}$-distribution. 
We do not fit an analytic function to $n(\mathbf{\zeta_{0}},z)$, as was done by \citet{rwh+12} and \citet{src+13}. 
We discuss the possible consequences of this for our results in \S3.3.3. 

We relate the dark matter halo virial velocities, $V_{\text{vir}}$, of galaxies in the G11 model to spheroid stellar velocity dispersions 
$\sigma$ \citep{bbh+03,mbb+08}. This assumption is discussed further in \S3.3.2. 
For each bin of $z$, $M_{1}$ and $M_{2}$, we find the average velocity dispersions of recently-merged 
galaxies in the G11 model hosting an SMBH of mass $M_{1}+M_{2}$. We use these values to specify $a_{H}$ for each bin of the 
discrete distribution $n(\mathbf{\zeta_{0}},z)$. 

\subsection{Evolution of binary SMBH orbits to the GW regime}

We assume that all SMBH binary orbits decay through interactions with fixed, isotropic, unbound cuspy stellar backgrounds, and 
through GW emission. The former scenario has been extensively studied numerically by \citet{q96}, \citet{shm06} and \citet{s10}. 
We assume a power-law stellar density distribution within the binary gravitational influence radius for all 
galaxies prior to mergers. For the majority of this paper, we additionally 
assume a stellar density profile power-law index of $\gamma=1.5$ corresponding to a mild stellar cusp \citep[see Equation 1 of][]{s10}. 
We consider variations in these assumptions further in \S3.3.2. 

We evolve the binary eccentricities, $e$, and semi-major axes, $a$, through scattering by unbound stars and loss of energy and angular 
momentum to GWs using expressions for $\frac{da}{dt_{r}}$ and $\frac{de}{dt_{r}}$  from Equations (15) and (16) of \citet{s10}. The effects of 
the ejection of stars that are \textit{bound} to the SMBHs \nocite{shm08}(Sesana, Haardt \& Madau 2008) are significant only for binary separations greater than 
$a_{h}$, and we hence neglect this phenomenon. 

We use the fits of \citet{shm06} for the rates of evolution of binary semi-major axes and eccentricities based on numerical scattering experiments 
(the `$H$' and `$K$' coefficients respectively from Tables 1 and 3 of Sesana et al. 2006). We log-interpolate the published 
values at binary component mass ratios of interest. As \citet{shm06} only provide rates of semi-major axis evolution for circular 
binaries, we assume here that the rate of semi-major axis evolution at a given semi-major axis is independent of eccentricity. This 
approximation leads to the semi-major axis evolution rate being underestimated by at most 20\% for the most eccentric binaries 
\citep[see Figure~3 of][]{shm06}. 
We also only use the seven values for our initial binary eccentricities (i.e., $e_{0}$) considered by \citet{shm06}; see their Table~3. 
These are 0, 0.15, 0.3, 0.45, 0.6, 0.75 and 0.9. 

By numerically integrating the expressions for $\frac{da}{dt_{r}}$ and $\frac{de}{dt_{r}}$ for each combination of $\mathbf{\zeta_{0}}$ and $z$, 
we first calculate the binary eccentricities, $e_{\text{GW}}$, at a rest-frame orbital frequency of $10^{-12}$\,Hz. Binaries with this orbital 
frequency emit negligible GW power in the PTA frequency band. The orbital frequency of a binary is given by 
\begin{equation}
f_{\text{orb}}=\frac{1}{2\pi}\left(\frac{G(M_{1}+M_{2})}{a^{3}}\right)^{1/2}.
\end{equation}
Letting $\mathbf{\zeta_{\text{GW}}}=[M_{1},M_{2},e_{\text{GW}}]$, we hence form the distribution function of binaries with orbital 
frequencies of $10^{-12}$\,Hz, $D_{\mathbf{\zeta_{\text{GW}}}}[N(\mathbf{\zeta_{\text{GW}}},z)]$, from the distribution of binaries at formation. 
If a binary at formation has $f_{\rm orb}>10^{-12}$\,Hz, we do not evolve the binary backwards in time to an orbital frequency of $10^{-12}$\,Hz.

To then specify the population of GW-emitting binary SMBHs, we need to calculate the numbers of binaries with different orbital frequencies. 
The GW luminosity, $L$, per unit frequency, $f_{r}$, of a binary SMBH depends on the masses $M_{1}$ and $M_{2}$, the eccentricity $e$, and the 
orbital frequency $f_{\text{orb}}$ \citep{pm63}.  
The functional form of $L(f_{r},\mathbf{\zeta})$ is given in, for example, Equation (2.6) of \citet{en07}. 
We now define a new parameter vector $\mathbf{\zeta}=[M_{1},M_{2},e,f_{\text{orb}}]$.

The distribution $D_{\mathbf{\zeta_{\text{GW}}}}[N(\mathbf{\zeta_{\text{GW}}},z)]$ can be used to specify the distribution function 
$D_{\mathbf{\zeta}}[N(\mathbf{\zeta},z)]$ using a continuity equation similar to Equation (35) of \citet{p01}:
\begin{equation}
\frac{d}{df_{\text{orb}}}\left[\frac{df_{\text{orb}}}{dt_{r}}D_{\mathbf{\zeta}}[N(\mathbf{\zeta},z)]\right]=-\frac{d}{dt_{r}}[D_{\mathbf{\zeta_{\text{GW}}}}[N(\mathbf{\zeta_{\text{GW}}},z)]]\delta(f_{\text{orb}}),
\end{equation}
where $\frac{d}{dt_{r}}[D_{\mathbf{\zeta_{0}}}[N(\mathbf{\zeta_{\text{GW}}},z)]]$ is the number of coalescences of binary SMBHs with parameters 
$\mathbf{\zeta_{\text{GW}}}$ per unit proper time $t_{r}$. The derivative $\frac{df_{\text{orb}}}{dt_{r}}$ is equivalent to 
$\frac{df_{\text{orb}}}{da}\frac{da}{dt_{r}}$, where $\frac{da}{dt_{r}}$ is given in Equation (15) of \citet{s10}. The solution is 
\begin{eqnarray}
D_{\mathbf{\zeta}}[N(\mathbf{\zeta},z)]&=&-\frac{d}{dt_{r}}[D_{\mathbf{\zeta_{\text{GW}}}}[N(\mathbf{\zeta_{\text{GW}}},z)]]\left(\frac{df_{\text{orb}}}{dt_{r}}\right)^{-1} \\
&=& -\frac{dz}{dt_{r}}\frac{d}{dz}[D_{\mathbf{\zeta_{\text{GW}}}}[N(\mathbf{\zeta_{\text{GW}}},z)]]\left(\frac{df_{\text{orb}}}{dt_{r}}\right)^{-1} 
\end{eqnarray}
We also associate each value of $f_{\text{orb}}$ with a unique value of $e$ by further integrating the expression for $\frac{de}{dt_{r}}$ from \citet{s10}.

Then, from Equation~(3), we have
\begin{equation}
\Omega_{\text{GW}}(f)=\int_{0}^{\infty}\left[\int...\int_{\mathbf{\zeta}}\frac{fL(f_{r})D_{\mathbf{\zeta}}[N(\mathbf{\zeta},z)]}{\rho_{c}c^{2}H(z)(1+z)} dM_{1}...df_{\text{orb}} \right]dz
\end{equation}
Recall that $f=f_{r}/(1+z)$. For consistency with other works, we calculate the characteristic strain spectrum, defined as 
\begin{equation}
h_{c}(f)=f^{-1}\left(\frac{3H_{0}^{2}}{2\pi^{2}}\Omega_{\text{GW}}(f)\right)^{1/2}.
\end{equation}

We perform the integral in Equation~(13) over $f_{\text{orb}}$ between $10^{-12}-10^{-5}$\,Hz. The upper orbital frequency limit corresponds to 
GW emission that is outside the PTA frequency band, even for binaries at high redshifts. For eccentric binaries, we consider radiation 
up to the 100th harmonic of $f_{\text{orb}}$ \citep{pm63}. 
We assume that binaries reach their last stable 
orbits at separations of three Schwarzschild radii of the more massive SMBH \citep{h02}, and neglect GW emission at smaller 
separations. 

\section{Predictions for the characteristic strain spectrum}

\subsection{Results}

As stated above, we consider four different initial eccentricity distributions: $w_{0}=0,\,0.1,\,0.35,\,0.93$. 
Recall that the $w_{0}=0$ case corresponds to 
all binaries being circular. Initially circular binaries are not expected to become eccentric \citep[e.g.,][]{s10}. The probability mass functions 
of the binary eccentricities, $e_{0}$, at $a_{H}$ in the three cases with $w_{0}>0$ are shown in Figure~1. For comparison, we also show in the bottom panel 
of Figure~1 a `thermal' probability mass function for $e_{0}$, derived from the probability density function $f_{e_{0}}=2e_{0}$ for $0\leq e_{0}\leq 1$. 
This would be expected if binary systems followed a purely Maxwell-Boltzmann distribution of energies \citep[e.g.,][]{a37}, as is roughly the case for galactic stellar 
binaries \citep{dm91}. 

\begin{figure}
\centering
\includegraphics[angle=-90,scale=0.7]{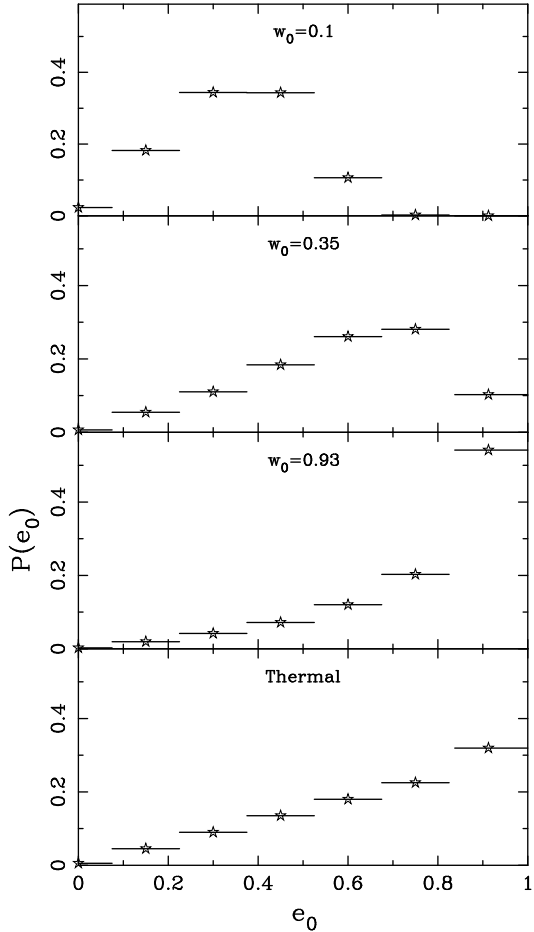}
\caption{Probabilities, $P(e_{0})$, of obtaining different values of $e_{0}$ (indicated by stars) for three initial eccentricity distributions 
defined by $w_{0}=0.1,\,0.35,\,0.93$ and Equation~(7) (top three panels), and for a thermal eccentricity distribution (bottom panel). 
The values of $e_{0}$ correspond to those considered by Sesana et al. (2006); see text for details.}
\end{figure}

\begin{figure}
\centering
\includegraphics[angle=-90,scale=0.35]{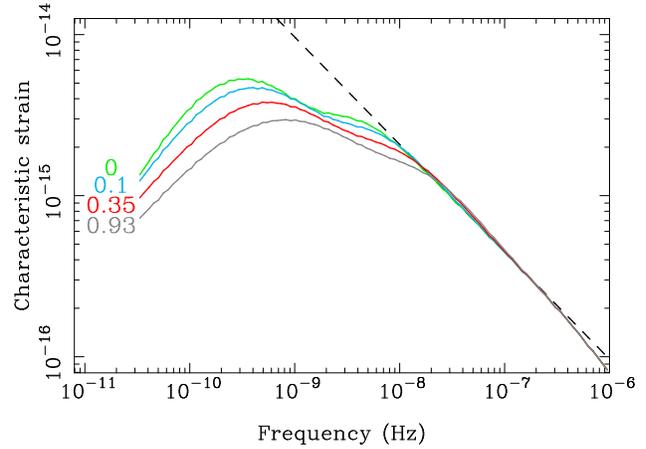}
\caption{The solid lines depict characteristic strain spectra for $w_{0}=0$ (green),  
$w_{0}=0.1$ (blue), $w_{0}=0.35$ (red) and $w_{0}=0.93$ (grey); the $w_{0}$ values for each line are given at the left of the plot. 
All curves were calculated assuming a stellar density profile index 
of $\gamma=1.5$. The black dashed line is the characteristic strain spectrum assuming circular orbits and purely GW-driven 
evolution for all SMBH binaries.}
\end{figure}

In Figure~2, we plot the characteristic strain spectra for each initial eccentricity distribution. Also depicted is the prediction in the circular 
(i.e., $w_{0}=0$), GW-driven case (i.e., for $\frac{da}{dt_{r}}$ including only GW-driven orbital decay for all $a$). This latter prediction 
corresponds to the standard $h_{c}(f)\propto f^{-2/3}$ power-law. In order to help highlight the physical effects at work, Figure~3 shows the 
characteristic strain spectra for each assumed $w_{0}$ contributed by binaries with combined masses in the ranges 
$10^{6.5}M_{\odot}-10^{10}M_{\odot}$ and $10^{10}M_{\odot}-10^{11}M_{\odot}$ respectively. 

The model we utilise for interactions between binaries and their stellar environments results in an 
attenuation of $h_{c}(f)$ in the PTA frequency band compared to the $f^{-2/3}$ power-law obtained in the circular, GW-driven case. 
For $w_{0}=0$, the signal is attenuated at frequencies $f\lesssim10^{-8}$\,Hz. At these frequencies, stellar interactions 
are the dominant binary orbital decay process, increasing $\frac{df_{\text{orb}}}{dt_{r}}$ in Equation~(12) and reducing 
the number of binaries observed per unit orbital frequency. For increasing $w_{0}$, the signal is further attenuated at 
low frequencies, although a slight ($\sim0.01$\,dex), increasing excess is present at frequencies between $10^{-8}$\,Hz and $10^{-7}$\,Hz. This is 
caused by two effects: eccentric binaries evolve faster than circular binaries, and eccentric binaries radiate GWs at 
higher harmonics of their orbital frequencies than circular binaries. 

The `substructure', or two bumps, in the characteristic strain spectra is a direct consequence of the mass-distribution of the binaries 
in our model. If $D_{\mathbf{\zeta_{0}}}[N(\mathbf{\zeta_{0}},z)]$ were smooth and analytic, the characteristic strain spectra 
would have only one clear peak. Here, however, we evaluate this distribution from the G11 semi-analytic model outputs 
(see Equation~(8)), which results in the distribution being incomplete at the high-mass end. These gaps in the distribution lead 
to the two apparent peaks in the characteristic strain spectra. 

As is evident in Figure~3, the first peaks of the spectra in Figure~2 are dominated by the highest-mass binaries, whereas the 
second peaks are dominated by lower-mass binaries. This is because the evolution of the highest-mass binaries begins to be 
GW-driven at lower frequencies than for less massive binaries. There are expected to be very 
few binaries in the (combined) mass range $10^{10}M_{\odot}-10^{11}M_{\odot}$; only $\sim50$ with $f_{\rm orb}\geq10^{-12}$\,Hz 
are expected to be present 
in the observable Universe according to the G11 model. In contrast, $\sim5\times10^{6}$ binaries are expected the the range 
$10^{6.5}M_{\odot}-10^{10}M_{\odot}$. The effects of sparsity in $D_{\mathbf{\zeta_{0}}}[N(\mathbf{\zeta_{0}},z)]$ are discussed 
further in \S3.3.3.

\begin{figure}
\centering
\includegraphics[angle=-90,scale=0.35]{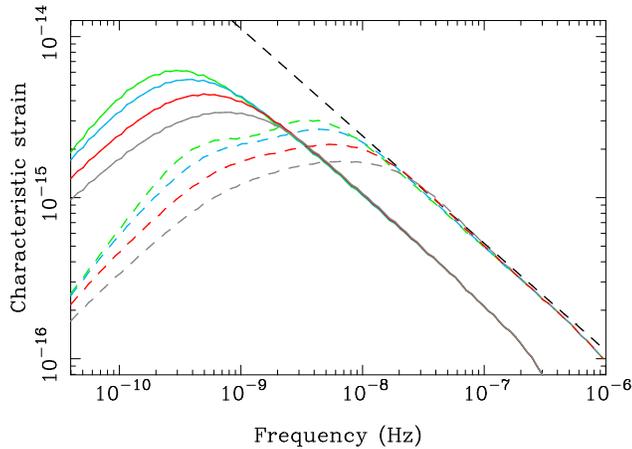}
\caption{Characteristic strain spectra contributed by binaries with total masses in the range $10^{6.5}M_{\odot}-10^{10}M_{\odot}$ (dashed curves) and in the range $10^{10}M_{\odot}-10^{11}M_{\odot}$ (solid curves). The colours represent different values of $w_{0}$ as in 
Figure~2; note that the orders of the low- and high-mass curves from top to bottom correspond to increasing $w_{0}$ as in Figure~2. 
We again show the characteristic strain spectrum for all SMBH binaries assuming circular orbits and purely GW-driven evolution as 
a black dashed line.}
\end{figure}

\subsection{Comparison with previous work}

Our results for the GW characteristic strain spectra from an eccentric binary SMBH population are broadly consistent with 
similar previous studies \citep{en07,s13}. Both these works find spectra which depart from the standard power law of the circular, GW-driven case 
at frequencies $f<10^{-8}$\,Hz. The results of \citet{s13} for binaries with eccentricities at formation of $0.7$ are in fact 
very similar to ours (see their Figure~2), with slight substructure evident along with the slight excess for $f>10^{-8}$\,Hz. 

Our results for $w_{0}=0$, however, differ somewhat from those of \citet{s13}. Whereas the maximum separation between 
the zero-eccentricity and high-eccentricity curves (the red solid and dashed curves in Figure~2 of Sesana 2013a) 
is approximately 0.5\,dex, the maximum difference between our curves for $w_{0}=0$ and $w_{0}=0.93$ in our Figure~2 
is 0.35\,dex. We also find similarly-shaped spectra for all $w_{0}$, whereas \citet{s13} has a clear single peak in their zero-eccentricity curve. 

The differences between our results and those of \citet{s13} for $w_{0}=0$ are caused by the nature of the respective 
binary SMBH mass distributions used. As discussed above, if the mass-distribution of binary SMBHs 
($D_{\mathbf{\zeta_{0}}}[N(\mathbf{\zeta_{0}},z)]$) were smooth and analytic, 
which is the case in \citet{s13}, only a single peak is expected. The reason for the similarity between our results and those of 
\citet{s13} for non-zero eccentricities may be because of some discreteness in the eccentric binary SMBH distribution used by 
\citet{s13}, as evidenced by the jagged nature of their strain spectrum at low frequencies.

The characteristic strain spectrum we predict in the circular, GW-driven case is $\sim0.15$\,dex lower than that predicted by 
\citet{src+13}. This difference is because we do not fit an analytic function to the discrete binary distribution $n(\mathbf{\zeta_{0}},z)$. 
We discuss this point further in \S3.3.3. 

\subsection{Uncertainties in the model predictions}

In this section, we describe the key uncertainties in our prediction of $h_{c}(f)$, which are summarised in Figure~4. 
We consider in turn the accuracy of the model predictions for SMBH demographics and coalescence rates and for the 
rate of evolution of binary systems, and the effects of incomplete high-mass binary SMBH distributions. 

\subsubsection{SMBH demographics and coalescence rates}

The merger rate of massive galaxies predicted by galaxy formation models \nocite{bdt07}(Bertone, De Lucia \& Thomas 2007) implemented in the 
Millennium simulation \citep{swj+05} has been shown to be consistent with observational estimates 
at redshifts $z<2$ \citep{bc09}. \citet{mbb+08} found that the model matches the observed quasar bolometric 
luminosity function at redshifts $z\leq1$ for a variety of assumed quasar lightcurves.  This, together with the reproduction of the local 
SMBH-galaxy scaling relations, suggests that the rate of formation of massive binary SMBHs at low redshifts is satisfactorily reproduced  
by the G11 semi-analytic model, which is used as the basis for this paper. 
Furthermore, the characteristic strain spectrum expected in the $w_{0}=0$ case for binaries with combined masses 
$M_{1}+M_{2}>2\times10^{8}M_{\odot}$ at redshifts $z\leq1$ has a maximum disparity with the unrestricted spectrum 
of 0.02\,dex. Hence, our model robustly predicts the contribution to the GW signal from massive, 
 low-redshift binaries, which are likely to dominate the total GW signal \citep[see also][]{wl03,mop12,s12}.
 
However, there remain a range of theoretical uncertainties. For example, the 
G11 model treatment of SMBHs does not include physically-motivated prescriptions for 
SMBH formation \citep[e.g.,][]{h13}, SMBH ejection caused by gravitational recoil following the coalescence of binary 
systems \citep[e.g.,][]{kon+13}, and does not account for any mass accreted onto SMBHs in merging galaxies prior to 
coalescence \citep[e.g.,][]{vvm+12}. 

There are also specific observational uncertainties in tuning the semi-analytic model. The current sample of SMBH and host galaxy bulge mass 
measurements, which is used to tune the quasar-mode SMBH accretion efficiency, allows for a $1\sigma$ confidence interval of $\sim0.2$\,dex in the 
SMBH masses \citep{src+13}. Similarly, the galaxy stellar mass function predicted by the G11 model is matched to Sloan Digital Sky Survey 
observations in the nearby Universe \citep[e.g.,][]{lw09}. 
These observations have a $\sim$0.2\,dex systematic uncertainty, with negligible contribution from 
cosmic variance \citep{lw09}, which corresponds (to first order) to a $\sim0.3$\,dex uncertainty in the galaxy merger rate. 

The uncertainty in SMBH masses corresponds to a $\sim0.3$\,dex uncertainty in $\Omega_{\text{GW}}(f)$, while the uncertainty in 
the merger rate translates directly to the range of predictions for $\Omega_{\text{GW}}(f)$ allowed by the observed galaxy stellar mass function. 
Combining both ranges results in a 0.4\,dex ($1\sigma$) uncertainty in $\Omega_{\text{GW}}(f)$, which corresponds to a 
0.2\,dex uncertainty in $h_{c}(f)$.

\subsubsection{The binary evolution model}

In this paper, we assume that all galaxies hosting SMBHs have spherically-symmetric central stellar density profiles 
that are power-law functions of radius, $r$, following \citet{s10}. That is, the stellar density, $\rho(r)$, is proportional to 
$r^{-\gamma}$, where we have hitherto  assumed $\gamma=1.5$. These profiles are equivalent to the central (asymptotic) behaviour 
of the \citet{d93} stellar potential and density models, which correspond well to high-resolution observations of the centres of 
nearby galaxy bulges \citep{fta+97}. Our assumption of a universal $\gamma$ is, however, not in agreement with observations, which typically show 
$1\lesssim\gamma\lesssim2$, with $\gamma=1$ corresponding to the most extreme `core' galaxies and $\gamma=2$ corresponding 
to the most extreme `power-law' galaxies \citep{d93,fta+97}. Furthermore, `core' galaxies are generally more massive, early-type 
systems with more massive SMBHs, and `power-law' galaxies are generally less massive, late-type systems with less massive SMBHs 
\citep[e.g.,][]{fta+97,mm13}. While we do not attempt to correlate $\gamma$ with galaxy properties from the G11 model, we 
show in Figure~4 characteristic strain spectra in the $w_{0}=0$ case for $\gamma=1$ and $\gamma=2$. The logarithmic 
differences between the spectra for these $\gamma$-values and the $w_{0}=0$ spectrum for $\gamma=1.5$ may be applied only approximately 
to the spectra for other $w_{0}$-values, because varying $\gamma$ varies both the rate of semi-major axis decay and the rate of eccentricity 
evolution for binaries.

The model that we use \citep{shm06,s10} for binaries evolving through separations less than $a_{H}$ due to 
interactions with fixed stellar backgrounds is qualitatively similar to the results of recent numerical simulations of 
dry (i.e., free of dynamically significant gas) galaxy merger remnants \citep{kpb+12}. However, as we show in Appendix~A, 
it is apparent that the model we use includes stronger stellar-driven orbital decay than the simulations of \citet{kpb+12}. This 
is despite our assumption (see \S2.2) that the rate of semi-major axis evolution is independent of binary eccentricity. This is 
not surprising, because the assumption of a fixed stellar background is qualitatively equivalent to the assumption of a full stellar loss-cone 
\citep[cf.][]{qh97,s10}. Hence, the model we use maximises binary orbital decay rates, in particular for spherically symmetric stellar distributions.

We are likely therefore to be overestimating the effects of stellar interactions on the binary SMBH population. The 
numerical simulations of \citet{kpb+12} suggest that the orbital frequencies at which binary SMBH evolution begins to be predominantly 
GW-driven are up to 0.45\,dex less than the corresponding frequencies that result from the model we use. This implies that the 
frequency below which the characteristic strain spectrum turns over from the $h_{c}(f)\propto f^{-2/3}$ power law may be 
up to 0.45\,dex lower than we predict. 

While the $V_{\text{vir}}-\sigma$ relation that we assume is established in the local Universe \citep{bbh+03}, it has not been 
studied at higher redshifts. Given the expected decrease in the stellar mass in a halo of a given mass with increasing redshift 
\citep{msm+10}, it is possible that we are overestimating the velocity dispersions of the stellar cores of merger remnants 
beyond the local Universe. This would imply that higher-redshift binaries decay more slowly than in our model, again increasing the 
low-frequency parts of the presented characteristic strain spectra. Further work is required to quantify the magnitude of this increase.  

Finally, while the assumed functional form of the $e_{0}$-distribution (Equation~(7)) is physically motivated, there may be some correlation 
between the orbital eccentricities of binaries with separations $a_{H}$, and their masses and redshifts. Additionally, a variety of 
studies find physical reasons for binaries to be quite circular upon formation \citep{cpv87,pr94,hfm03}, which suggests that low-$w_{0}$ 
values may be preferred. Current numerical simulations \citep[e.g.,][]{kpb+12} have not been run with a sufficient range of initial 
conditions to provide conclusive results on this point. 

\subsubsection{Accounting for discreteness in the binary SMBH distributions from the G11 model}

Given the distribution $n(\mathbf{\zeta_{0}},z)$ (Equation (8), \S2.2) of binary SMBHs, we have examined the \textit{expected value} of the 
GW characteristic strain spectrum. However, the binary SMBH distribution that we use is not exactly the  
distribution expected from the G11 semi-analytic model implemented within the 
Millennium simulation. The Millennium simulation provides a single realisation of the dark matter halo merger history within a large 
comoving volume, and the G11 prescriptions specify properties of the galaxies, and SMBHs, associated with the halos. To form the 
discrete binary distribution 
$n(\mathbf{\zeta_{0}},z)$, we count the numbers of binary SMBHs forming in the entire Millennium volume between redshift snapshots in 
bins of $M_{1}$ and $M_{2}$, assigning values of $e_{0}$ to each binary according to our $e_{0}$-distribution. However, despite the large 
volume, $n(\mathbf{\zeta_{0}},z)$ is poorly populated for high $M_{1}$ and $M_{2}$ at every redshift. To estimate the \textit{expected} nature of this 
distribution, statistical modelling is required. This was done by \citet{src+13}, who considered the circular, GW-driven case, and found that the modelling  
resulted in a characteristic strain spectrum increased by 0.15\,dex. However, the distribution $n(\mathbf{\zeta_{0}},z)$ has two more dimensions 
(an extra mass dimension and the eccentricity dimension) than that considered by \citet{src+13}, which significantly complicates the modelling. 
Instead, we simply consider it possible that the strain spectrum we have derived may be up to 0.15\,dex larger. 

A qualitatively similar effect was pointed out by \nocite{svc08}(Sesana, Vecchio \& Colacino 2008), who compared characteristic strain spectra generated from realisations of the 
binary SMBH population of the Universe to the spectrum expected on average, in the circular, GW-driven case. Whereas the average spectrum 
was a power-law proportional to $f^{-2/3}$, individual realisations had a lower amplitude at higher frequencies. This was because 
the numbers of binaries radiating GWs at a given frequency (per unit frequency) decreases with increasing frequency, implying that, for example, 
there is a frequency above which the expected number of sources is less than unity. However, the correct model for the \textit{average} characteristic 
strain spectrum still had $h_{c}(f)\propto f^{-2/3}$ for all $f$, despite all realisations of the spectrum being below this power-law at high frequencies. 
This situation is analogous to our suggestion of an increase in the characteristic strain spectrum if the average 
behaviour of $n(\mathbf{\zeta_{0}},z)$ were correctly modelled.

We also do not attempt here to describe the statistical nature of the GW signal, as was done by  \citet{rwh+12} in the circular, GW-driven case. 
\citet{rwh+12} modelled a GW signal that was mildly non-Gaussian, with individual sources dominant at all GW frequencies of interest to PTAs. 
\citet{src+13} further showed that assuming non-Gaussian statistics for the GW signal caused constraints on $\Omega_{\rm GW}$ to degrade 
by $\sim20\%$. This reflects the fact that realisations of $\Omega_{\rm GW}(f)$ at a particular frequency $f$ would have a larger variance in the 
non-Gaussian case than in the Gaussian case.

As discussed in \S3.1, environment-driven binary evolution causes the highest-mass binaries to dominate $\Omega_{\rm GW}(f)$ at 
low frequencies to a greater extent than in the purely GW-driven case. This, coupled with the sparsity of these binaries in our calculations, causes 
the low-frequency substructure in the characteristic strain spectra for all $w_{0}$ in Figure~2. Our results, however, suggest a more general conclusion: 
that, at low frequencies, environment-driven binary evolution causes the variance in realisations of $\Omega_{\rm GW}(f)$ to be significantly increased 
relative to the assumption of only GW-driven evolution. Including this increased variance in $\Omega_{\rm GW}(f)$ at low frequencies in 
the calculation of PTA upper bounds on $\Omega_{\rm GW}(f)$ \citep[e.g.,][]{src+13} would cause these constraints to be further degraded 
relative to constraints based on the work of \citet{rwh+12}.

\subsubsection{Synthesis of uncertainties in $h_{c}(f)$}

\begin{figure}
\centering
\includegraphics[angle=-90,scale=0.35]{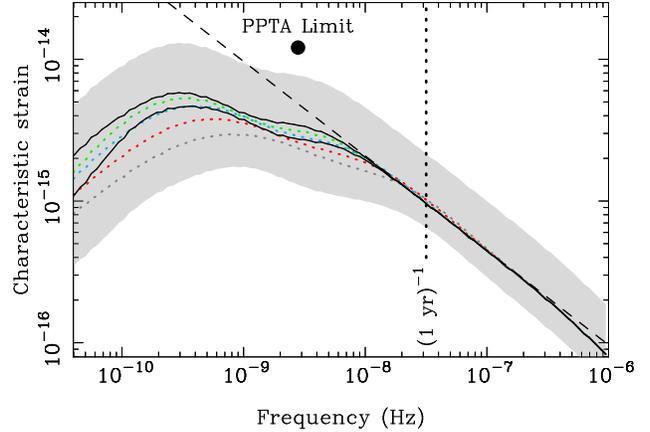}
\caption{The four coloured, dotted curves are the characteristic strain spectra for the four $w_{0}$ cases we consider, also shown with the 
same colours in Figure~2. The upper solid black curve corresponds to a stellar density profile index of  $\gamma=1$, and the lower solid 
black curve corresponds to $\gamma=2$; both are calculated assuming $w_{0}=0$, and so may be compared with the green 
(uppermost, $w_{0}=0$)
dotted curve. The grey shaded area represents 
an approximate 68\% confidence interval in our prediction of $h_{c}(f)$, given observational errors in the SMBH-bulge mass relation and the galaxy 
stellar mass function (a 0.4\,dex range), allowing for the full range of $w_{0}$ values, and including a possible increase of 0.15\,dex 
in the predictions if the binary SMBH population statistics were accurately specified \citep{src+13}. 
The black dot indicates the most recent 95\% confidence upper limit on the stochastic Gaussian GW signal \citep[][see text for details]{src+13}. 
The characteristic strain spectrum calculated here in the circular, GW-driven case (and shown in Figures 2 and 3) is displayed as a 
black dashed line. The vertical dotted line indicates a frequency of $(1\,{\rm yr})^{-1}$.}
\end{figure}

We refer the reader to Figure~4, where we show an approximate $1\sigma$ confidence interval on the characteristic strain spectrum according to 
the model we describe. This interval represents our uncertainty in the expected value of the signal, not the realisation-to-realisation 
uncertainty. The interval encompasses the maximum possible ranges of $w_{0}$ and $\gamma$ (see \S3.3.2), and also includes observational 
uncertainties in the SMBH-bulge mass relation and in the galaxy stellar mass function (see \S3.3.1). We also include our assertion that the characteristic 
strain spectrum could be up to 0.15\,dex larger than what we calculate if the binary SMBH distribution were correctly specified (see \S3.3.3).

It is clear that that there is relatively more uncertainty in our prediction at frequencies $f\lesssim2\times10^{-8}$\,Hz, where environmental interactions and binary eccentricities may affect the signal. We have also not included our uncertainty in the specific model for environment-driven binary SMBH
evolution. As discussed in \S3.3.2, the model we use may represent the maximum level of binary-environment coupling; other models may result 
in the characteristic strain spectrum being boosted at low frequencies relative to our prediction. For example, the model of \citet{kpb+12} 
suggests that the effects of environmental interactions may only be relevant for $f\lesssim7\times10^{-9}$\,Hz (also see Appendix~A). 
We have also weighted each $w_{0}$-value equally, whereas it is possible that low-$w_{0}$ values are preferred over high-$w_{0}$ values.

In Figure~4, we also indicate the best upper bound on a stochastic, Gaussian GW background from binary SMBHs, published recently 
by the Parkes Pulsar Timing Array \citep[PPTA;][]{src+13}. This upper bound corresponds to 
$\Omega_{\rm GW}(2.8\,{\rm nHz})<1.3\times10^{-9}$ with 95\% confidence. While PTA bounds are traditionally shown as 
wedges \citep[e.g.,][Figure~13]{svc08} on characteristic strain spectrum plots, \citet{src+13} argued that their limit was applicable only at a single 
GW frequency. We hence display this limit as a single dot. 

Our prediction for the characteristic strain spectrum at a frequency of $f=(1\,{\rm yr})^{-1}$ of $6.5\times10^{-16}<h_{c}<2.1\times10^{-15}$ 
(with approximately 68\% confidence) is broadly consistent with previous results \citep[e.g.,][]{wl03,svc08,s12,src+13} that considered the circular, 
GW-driven case. Indeed, for $f\gtrsim2\times10^{-8}$\,Hz, the predicted characteristic strain spectrum closely resembles the 
$h_{c}(f)\propto f^{-2/3}$ power law expected in the circular, GW-driven case, with the exception that for larger $w_{0}$ slightly more 
signal is present. The departure from the $f^{-2/3}$ power law at $f\sim3\times10^{-7}$\,Hz is caused by binary SMBHs radiating at these 
frequencies reaching their last stable orbits and not being included in our calculations \citep[see, e.g.,][]{wl03}.  

\subsection{Summary of PTA implications}

Our results suggest a challenging future for attempts at detecting the GW background from binary SMBHs with PTAs. 
The frequency of optimal sensitivity for PTAs generally corresponds to the inverse of the characteristic observation time \citep[e.g.,][]{src+13}. 
Typical observation times of $5-30$\,yr \citep{mhb+13} imply that the properties of the GW signal at frequencies in the range $5\times10^{-10}$\,Hz 
to $10^{-8}$\,Hz are of primary importance for PTA work. The model we use in this paper implies that the GW characteristic strain spectrum 
may be reduced throughout this frequency range relative to the circular, GW-driven case (i.e., with respect to a $h_{c}(f)\propto f^{-2/3}$ power law). 
For example, \citet{src+13} 
presented a single-frequency constraint on $\Omega_{\text{GW}}(f)$ that is inconsistent with a variety of astrophysical predictions 
assuming circular, GW-driven binaries. However, the constraint is at a frequency where the characteristic strain spectrum we predict 
(see Figure~4) is reduced by at least 0.08\,dex relative to the circular, GW-driven case. 
More generally, the gains in sensitivity to a GW background with observing time, estimated assuming $h_{c}(f)\propto f^{-2/3}$ \citep[e.g.,][]{sej+13}, 
are likely to be overestimated. Furthermore, it is possible that at low frequencies the increased contribution to the total GW signal 
of rare, massive binary SMBHs relative to the circular, GW case will cause the signal at these frequencies to be more non-Gaussian than suggested 
by \citet{rwh+12}. 

However, our results require significant refinement. It is clear from Figure~4 that our prediction for the characteristic strain spectrum at low frequencies 
is quite uncertain. Besides this uncertainty, the model we use in this paper for the coupling between binary SMBHs and stellar environments 
\citep{shm06,s10} may in fact maximise the strength of this coupling (see \S3.3.2 and Appendix~A). Also, it is possible that 
lower-eccentricity scenarios may be preferred over the higher-eccentricity scenarios. Both the above possibilities would result in the 
low-frequency parts of the characteristic strain spectrum being increased relative to our predictions. We strongly urge further work on modelling 
the evolution of binary SMBH orbits in a variety of realistic galaxy merger scenarios. This is of significant importance for predicting the 
strength of the GW signal from binary SMBHs in the PTA frequency band.

\section{Predictions for GW bursts}

\subsection{The distribution of GW bursts}

The prospect of detecting GW bursts with PTAs has been pursued recently by a number of authors 
\citep[e.g.,][]{fl10,p12}. GW burst detection algorithms generally contain few assumptions about the source properties, 
except that they search for a strong signal confined to a short time-period. Here, we focus on the properties of GW bursts 
from eccentric binary SMBHs, and use our distribution of binary SMBHs, $D_{\mathbf{\zeta}}[N(\mathbf{\zeta},z)]$, to predict 
the distribution of burst events. 

\begin{figure}
\centering
\includegraphics[angle=-90,scale=0.7]{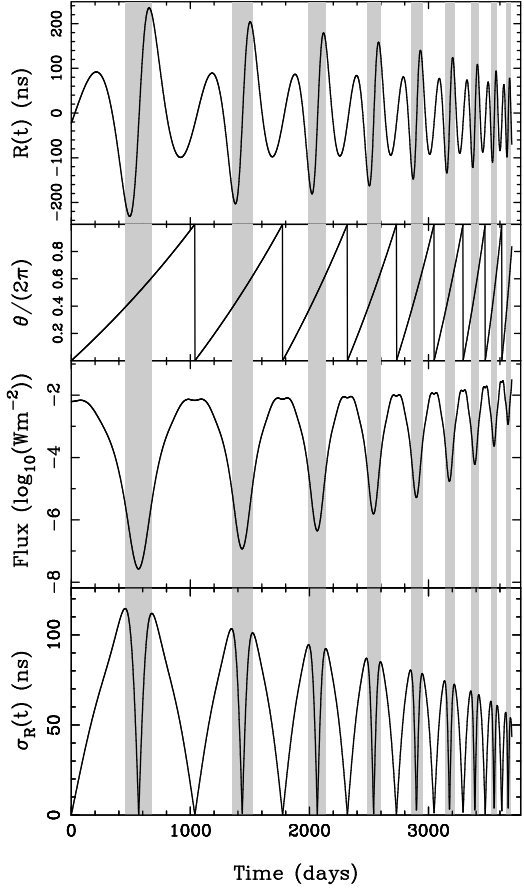}
\caption{Properties of GW signal and induced ToA variations for a binary SMBH with eccentricity 0.8, orbital period 
(at the Earth) of 3.1\,yr at the starting time, component masses $M_{1}=10^{10}\,M_{\odot}$ and $M_{2}=5\times10^{9}\,M_{\odot}$, 
and redshift 0.1. The grey shadings in each panel represent the time-periods identified as GW bursts. 
The panels, from the top, are described in turn. \textit{First panel:} the ToA variations corresponding to the metric 
perturbations at the Earth, for arbitrary orientation parameters, with the mean subtracted. 
\textit{Second panel}: the orbital phase $\theta$, measured from the common line of nodes and periapse. \textit{Third panel}: the GW flux at the Earth. 
\textit{Fourth panel}: the rms induced pulsar ToA variations, averaged over all orientation parameters. The time coordinate 
is measured at the Earth. Not all minima in the bottom curve are at a value of zero because of imperfect numerical 
sampling.}
\end{figure}

It is necessary to form a definition of a GW burst from an eccentric binary SMBH in terms 
pulsar timing data products. Pulsar timing is the practice of measuring the times of arrival (ToAs) of pulses from 
millisecond radio pulsars and fitting a physical model to these measurements.
In Appendix~B, we describe how GWs from eccentric binaries affect pulsar timing measurements by 
inducing variations to ToAs. We present an expression for the rms deviation, $\sigma_{R}(t)$, of the ToA variations as a 
function of time caused by a given binary SMBH in Equation~B11, averaged over all orientation parameters. 

In Figure~5, we show the orbital phase, $\theta$, the expected energy flux in GWs at the Earth and $\sigma_{R}(t)$ 
as functions of time for a binary SMBH with eccentricity 0.8 and orbital period at the Earth of 3.1\,yr at the starting time, 
component masses $M_{1}=10^{10}\,M_{\odot}$ and $M_{2}=5\times10^{9}\,M_{\odot}$, and redshift 0.1. We also show the 
induced ToA variations corresponding to the GW metric perturbation at the Earth, $R(t)$, for arbitrary orientation parameters 
(binary inclination $i=1$\,rad, line of nodes orientation $\phi=0.5$\,rad). The 
orbital evolution of the binary was calculated using the work of \citet{pm63}, and the energy flux at the Earth is 
averaged over binary inclination. The time-intervals considered to be GW bursts are highlighted in all panels of 
Figure~5. These `bursts' correspond to the times of the largest change in the shortest amount of time in the ToA variations 
(see the top panel of Figure~5), and can be identified using $\sigma_{R}(t)$. 
We define the true burst amplitude, $R_{\text{burst}}$, to be twice the peak value of $\sigma_{R}(t)$, because that represents 
the expected peak-to-peak variation for a burst. The burst duration, $T_{\text{burst}}$, is the time-interval between peaks in $\sigma_{R}(t)$, represented by the widths of the shaded intervals in Figure~5. 
It is interesting that the GW bursts in the ToAs correspond to the motion of the binary through apastron, rather 
than periastron. This is because the GW-induced ToA variations are given by the time-integral of the GW amplitude 
as a function of time, as outlined in Appendix~B. 

The qualitative properties of $\sigma_{R}(t)$ in Figure~5 apply to binaries with any component masses, orbital period and 
eccentricity. That is, there are two peaks per rotation period, separated by less in orbital phase for more eccentric binaries, and 
separated by half an orbital phase for circular binaries. For a binary specified by $\mathbf{\zeta}$ and $z$, we integrate 
the equations for the evolution of the orbit \citep{pm63} from zero orbital phase to numerically calculate $R_{\text{burst}}$ 
as the mean of the first two peaks in $\sigma_{R}(t)$, and $T_{\text{burst}}$ as the time-interval between peaks. 

We use the distribution of binary SMBHs, $D_{\mathbf{\zeta}}[N(\mathbf{\zeta},z)]$, to calculate the distribution 
of GW bursts. As described above, this distribution is specified as the number of binaries in discrete bins of width 
$\Delta M_{1}$, $\Delta M_{2}$, $\Delta z$, $\Delta f_{\text{orb}}$ and $\Delta e_{\text{GW}}$, where the eccentricity 
bin-widths depend on the other parameters. Scaling this distribution by the comoving volume shell between redshifts 
$z-\Delta z/2$ and $z+\Delta z/2$ specifies the number of observable binary SMBHs. For parameters at the 
midpoints of each bin, we calculate $R_{\text{burst}}$ and $T_{\text{burst}}$. We approximate the burst rate 
from binaries in a bin as the number of binaries divided by their period observed at the Earth, and 
record the expected number of bursts in a 10\,yr time-span. 

\begin{figure*}
\centering
\includegraphics[angle=-90]{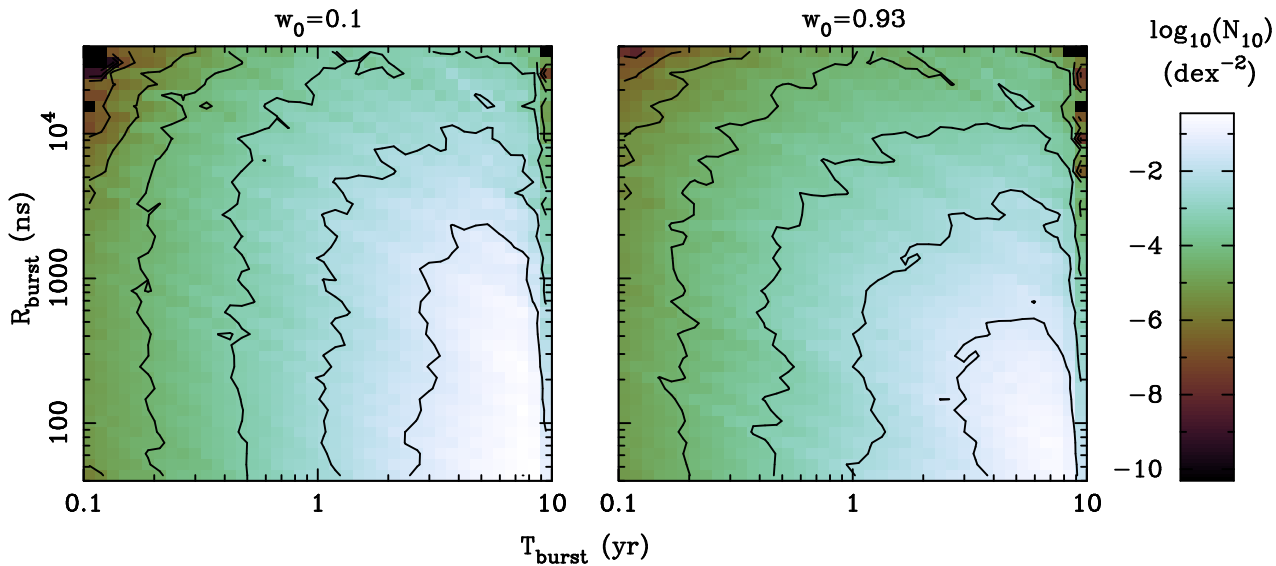}
\caption{Illustrations of the distributions of GW bursts in the $w_{0}=0.1$ (left) and $w_{0}=0.93$ (right) cases. The shading represents the 
expected number of bursts in a 10\,yr time-span, $N_{\rm 10}$, per dex$^{2}$. The distributions are binned over 0.05\,dex in duration 
and 0.075\,dex in amplitude. The contours connect regions at intervals of factors of 10 below the peak.}
\end{figure*}

\subsection{Results}

Using the distributions of binary SMBHs for $w_{0}=0.1$ and $w_{0}=0.93$, we calculated the numbers of GW 
bursts for different values of the expected maximum level of ToA variations, $R_{\text{burst}}$, and the 
duration, $T_{\text{burst}}$. We depict the distributions of GW bursts in 
Figure~6 as the number of bursts per 10\,yr observation time, $N_{\rm 10}$, per dex$^{2}$, in bins of 0.075\,dex in $R_{\text{burst}}$ and 
0.05\,dex in  $T_{\text{burst}}$. We only considered bursts with $R_{\text{burst}}\geq40$\,ns and 
0.1\,yr\,$\leq T_{\text{burst}}\leq$\,10\,yr. An rms ToA variation of 40\,ns corresponds to the best timing precisions 
currently achieved for millisecond radio pulsars \citep{ovd+13,george13}. 

In total, we predict 0.06 bursts per 10\,yr in the $w_{0}=0.93$ case with these strengths and durations, as compared 
with 0.12 bursts per 10\,yr in the $w_{0}=0.1$ case. This difference in the total number of bursts is because 
of the smaller number of binary SMBHs that we expect to observe if the population is generally more eccentric. However, 
we note that bursts from low-eccentricity binaries, which will dominate the burst population in the $w_{0}=0.1$ case, 
may be less detectable than bursts from high-eccentricity binaries. There are proportionally more short-duration bursts 
in the $w_{0}=0.93$ case than in the $w_{0}=0.1$ case, because larger binary eccentricities result in shorter bursts. 

In both cases, the burst distribution is quite 
heavily skewed towards long bursts, with approximately a factor of 100 more bursts expected with $\sim8$\,yr durations 
than with $\sim1$\,yr durations. There are also fewer bursts with durations longer than $\sim8$\,yr in 
both cases. This typical burst duration corresponds to binaries with separations where GW-driven evolution is 
equivalent to evolution driven by stellar environments. 

The typical combined masses of the binary SMBHs that produce GW bursts are $\sim10^{10}\,M_{\odot}$. In Figure~7, 
we show the distributions of the combined masses of all binary SMBHs producing the bursts in the distributions 
in Figure~6. The distributions are similar in shape, although the distribution for $w_{0}=0.93$ includes relatively more 
high-mass binaries than the distribution for $w_{0}=0.1$. This is because, in the $w_{0}=0.93$ case, lower-mass binaries 
are less likely to be able to produce strong GW bursts because they are likely to be more eccentric. More eccentric binaries 
of a given mass and orbital period produce typically weaker bursts (see Equation~(B11) in Appendix~B). 

\begin{figure}
\centering
\includegraphics[angle=-90,scale=0.35]{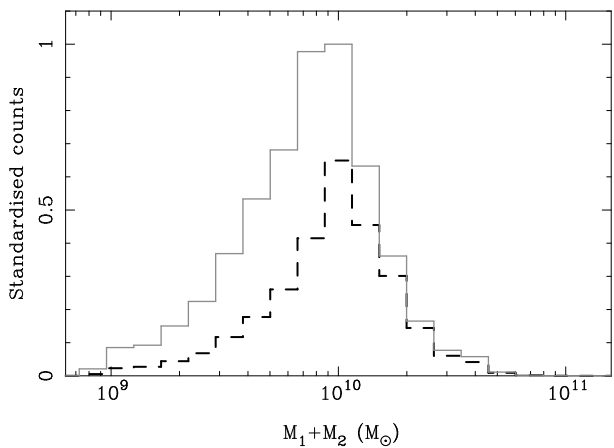}
\caption{The distributions of combined masses ($M_{1}+M_{2}$) of binary SMBHs contributing to the GW burst distributions 
presented in Figure~6. The solid grey histogram corresponds to the $w_{0}=0.1$ case, and the dashed black histogram 
corresponds to the $w_{0}=0.93$ case. Both histograms are normalised to the peak of the $w_{0}=0.1$ histogram.}
\end{figure}

In summary, we find:
\begin{enumerate}
\item For bursts with durations between 0.1\,yr and 10\,yr, and with expected maximum ToA variations of $>40$\,ns, 
we predict between 0.06 and 0.12 bursts per 10\,yr observation, with lower burst rates corresponding to 
higher-eccentricity binary SMBH populations.
\item Higher-eccentricity binary populations result in relatively more shorter duration bursts than lower-eccentricity 
populations. 
\item However, the burst rate decreases by a factor of 10 per $\sim0.4$\,dex below a duration of $\sim8$\,yr. 
This also appears to be the most likely duration, with few bursts longer than 8\,yr expected. 
\item The burst rate decreases by a factor of 10 per $\sim0.8$\,dex increase in amplitude. 
\end{enumerate}

Various uncertainties discussed in \S3.3 also apply to these calculations. 
The uncertainty in the galaxy merger rate will also directly correspond to the uncertainty in the GW burst rate (i.e., 0.3\,dex). 
Given that the high-end power-law logarithmic slope of the SMBH mass function in the G11 model is $\sim-2$, the 0.2\,dex uncertainty in the 
SMBH masses will, to first order, correspond to an uncertainty of 
0.4\,dex in the merger rate. Therefore, the $1\sigma$ uncertainty in the burst rate from the model is approximately 0.5\,dex. 

To our knowledge, only one study has attempted to predict the properties of GW bursts from a population of 
eccentric binary SMBHs \citep{lzz+12}. While our modelling methods and definition of GW bursts differ substantially 
from this work, we agree with these authors that it is unlikely that current PTAs will be able to detect GW bursts from binary 
SMBHs. The rarity of short GW bursts (lasting around 1\,yr) from binary SMBHs suggests that high-cadence PTA observations 
targeting such bursts are not well motivated. 

\section{Summary of results}

We have used a semi-analytic model for galaxy and SMBH formation and evolution \citep{gwb+11} implemented in the Millennium simulation 
\citep{swj+05}, augmented with a model for the evolution of binary SMBHs within fixed stellar backgrounds \citep{shm06,s10}, to predict the 
properties of low-frequency GWs from binary SMBHs. We specify the form of a 
phenomenological distribution of initial binary eccentricities, and consider a selection of cases with differing levels of typical binary eccentricity. 

Our quantitative results are uncertain due to a variety of factors. The range of initial binary eccentricity distributions that we 
consider corresponds to a 0.4\,dex variation in the characteristic strain spectrum at low frequencies. Moreover, uncertainties 
in the tuning of the G11 model provide another 0.2\,dex of uncertainty in the spectrum at all frequencies. There is also uncertainty in our estimate of 
the binary SMBH distribution predicted by the G11 model, in particular for the most massive binaries. Finally, while the G11 model is likely to 
provide a satisfactory representation of the merger rate of massive, low-redshift galaxies, the binary evolution model that we use 
may overestimate binary hardening caused by stellar interactions.

Our specific findings are as follows:
\begin{enumerate} 

\item The expected characteristic strain spectrum of the GW background from binary SMBHs will turn over from the standard 
$h_{c}(f)\propto f^{-2/3}$ power law at a frequency up to $2\times10^{-8}$\,Hz. The turn-over frequency 
depends on the efficiency of stellar interactions in extracting energy and angular momentum from binary SMBHs, as well as 
the typical binary eccentricities at formation.

\item The nature of the spectrum at frequencies below the turn-over frequency is extremely uncertain, and depends 
on the numbers of massive ($M_{1}+M_{2}\geq10^{10}M_{\odot}$) binaries and on binary eccentricities. 
The most massive binary SMBHs predominantly produce the lowest-frequency parts of the spectrum, and their numbers depend 
strongly on the strength of their coupling to their environments. The spectrum will be attenuated if binaries with typically 
larger eccentricities are present.

\item The most massive eccentric binaries will very rarely produce GW bursts detectable in pulsar 
timing data. A larger-eccentricity binary population will produce fewer bursts that are typically shorter and weaker. 
Our results suggest that GW bursts from binary SMBHs do not provide viable targets for PTA observations. 

\end{enumerate}

We emphasise a set of key implications of our work for PTAs:
\begin{enumerate}

\item Given the expected low-frequency turn-over in the GW characteristic strain spectrum, along with the large uncertainty in the 
signal at these frequencies, the increase with time of PTA sensitivities to a GW background from binary SMBHs will not be as 
strong as currently thought.

\item Short-duration, strong GW bursts from eccentric binary SMBHs are 
unlikely to occur during typical PTA dataset lengths. 

\item PTA data analysts cannot assume $h_{c}(f)\propto f^{-2/3}$ when searching for a GW background from 
binary SMBHs. Indeed, \textit{model-independent} searches cannot assume any particular spectral shape. 

\item \textit{Model-dependent} searches and constraints need to carefully account for the uncertainty in model predictions.

\end{enumerate}

\section{Conclusions}

In this paper, we predict both the GW background characteristic strain spectrum and the distribution of strong GW bursts from eccentric binaries. At a GW frequency of (1\,yr)$^{-1}$, we predict a characteristic strain of 
$6.5\times10^{-16}<h_{c}<2.1\times10^{-15}$ with approximately 68\% confidence. 

Accelerated binary evolution driven by three-body stellar interactions causes the characteristic strain spectrum to be diminished with 
respect to a $h_{c}(f)\propto f^{-2/3}$ power-law at $f\lesssim2\times10^{-8}$\,Hz. At these low frequencies, the signal is 
further attenuated 
if binary SMBHs are typically more eccentric at formation. The low-frequency signal may be dominated by a few 
binaries with combined masses ($M_{1}+M_{2}$) greater than $10^{10}\,M_{\odot}$, to a larger extent than predicted in the circular, GW-driven case 
\citep{rwh+12}. Numerous uncertainties, however, affect our results. These include observational uncertainties in parameters of our model, and 
theoretical uncertainties in the efficiency of coupling between binary SMBHs and their environments. 

We also expect between 0.06 and 0.12 GW bursts that produce $>$40\,ns amplitude ToA variations over a 10\,yr observation 
time. Larger typical binary eccentricities at formation will result in fewer events than if binaries are less eccentric at formation. 
These bursts are caused by binary SMBHs with combined masses of $\sim10^{10}\,M_{\odot}$, and typically last 
$\sim8$\,yr. Shorter, stronger bursts are significantly less likely, as are longer bursts.

Upcoming radio telescopes with extremely large collecting areas, such as the Five hundred metre Aperture Spherical Telescope \nocite{lnp13}(FAST, Li, Nan \& Pan 2013) and the Square Kilometre Array \citep[SKA,][]{ckl+04} are likely to significantly expand the sample of pulsars with sufficient timing precision for GW 
detection as compared to current instruments. 
PTAs formed with FAST and the SKA will hence be sensitive to a stochastic GW signal at much higher frequencies than current PTAs, 
which is desirable given the results we present. 

The mechanism by which binary SMBHs are driven to the GW-dominated regime must involve some form of binary-environment coupling. 
Hence, independent of the exact model, 
\textit{there will always be some low-frequency attenuation of the GW signal relative to the circular, GW-driven binary case.} 
Our results indicate that this attenuation occurs within the PTA frequency band. 
However, the strength of the binary-environment coupling is quite uncertain, and we urge future work on this topic.  
Finally, as also emphasised in previous works \citep{en07,s13}, constraining or measuring the spectrum of the GW background at a 
number of frequencies would provide an excellent test of models for the binary SMBH population of the Universe.

\section*{Acknowledgements}

The authors thank Alberto Sesana, Sarah Burke-Spolaor, Yuri Levin, Simon Mutch and Jonathan Khoo for useful discussions. 
V.R. is a recipient of a John Stocker Postgraduate Scholarship from the 
Science and Industry Endowment Fund, and J.S.B.W. acknowledges an Australian Research Council Laureate Fellowship. The Millennium 
and Millennium-II Simulation databases used in this paper and the web application providing online access to them were constructed as 
part of the activities of the German Astrophysical Virtual Observatory. This work was performed on the swinSTAR supercomputer at the 
Swinburne University of Technology. 

\appendix

\section{Testing our implementation of a binary SMBH evolution model}

In the main text, we use the results of \citet{shm06} and \citet{s10} (hereafter collectively referred to as S06) to model the evolution 
of binary SMBHs in fixed stellar backgrounds for separations less than $a_{H}$ (see Equation~(6)). S06 numerically solved 
three-body scattering problems for binary SMBHs interacting with stars on radial, intersecting orbits drawn from a spherically-symmetric, 
fixed distribution, and provided fitting formulae for the binary hardening and eccentricity growth rates as functions of binary properties. 
We use these fitting formulae to evolve binary SMBH orbits as described in \S2.3. 

Here, we compare this method of evolving binary SMBHs with recent $N$-body simulations of binary SMBH evolution in 
merging galaxies of various mass ratios and stellar density distributions \citep[][hereafter K12]{kpb+12}. K12 simulated the 
mergers of spherical galaxies with various mass ratios, power-law stellar cusp density profiles with various indices, with 
typical approach trajectories from cosmological simulations. The SMBHs were traced until separations close to, and in some 
cases beyond, where the GW-driven orbital decay dominated the orbital decay caused by three-body stellar interactions. 
By extrapolating the binary orbits assuming constant stellar-driven hardening rates and eccentricities, K12 estimated 
binary semi-major axes, $a_{\text{K12}}$, below which GW-driven evolution dominated. The accuracy of these 
extrapolations was confirmed using a selection of simulations including post-Newtonian corrections to the binary SMBH orbits. 

Here, we take the final eccentricities and semi-major axes of the binaries in each of the scenarios considered by 
K12 for which $a_{\text{K12}}$ was estimated, and evolve the binaries using the binary evolution model of S06 to 
estimate an equivalent quantity to $a_{\text{K12}}$, $a_{\text{S06}}$. We list the ratios $a_{\text{S06}}/a_{\text{K12}}$ 
in Table~A1 both without (column 4) and with the binary eccentricity held fixed (column 5) for each 
relevant scenario of K12. The cusp density profile indices, $\gamma$, and the galaxy and SMBH mass ratios, $q$, 
are given in columns 2 and 3 respectively.

\begin{table}
\centering
\caption{Comparison between decoupling times for S06 and K12 models.}
\begin{tabular}{ccccc}
\hline
Model & $\gamma$ & $q$ & $\frac{a_{\text{S06}}}{a_{\text{K12}}}$ & $\frac{a_{\text{S06}}}{a_{\text{K12}}}$ (fixed $e$) \\
\hline
A1 & 0.5 & 0.1 & 0.51 & 0.47 \\
A2 & 0.5 & 0.25 & 0.72 & 0.53 \\
A3 & 0.5 & 0.5 & 0.62 & 0.57 \\
A4 & 0.5 & 1.0 & 0.68 & 0.65 \\
B1 & 1.0 & 0.1 & 0.58 & 0.36 \\
B2 & 1.0 & 0.1 & 0.74 & 0.38 \\
B3 & 1.0 & 0.1 & 0.87 & 0.41 \\
B4 & 1.0 & 0.1 & 0.93 & 0.47 \\
D1 & 1.75 & 0.1 & 0.72 & 0.41 \\
D2 & 1.75 & 0.1 & 0.68 & 0.44 \\
D3 & 1.75 & 0.1 & 0.62 & 0.48 \\
D4 & 1.75 & 0.1 & 0.63 & 0.51 \\
\hline 
\end{tabular}
\end{table}

We find $0.1<\frac{a_{\text{S06}}}{a_{\text{K12}}}<1$ in all cases. This implies that the S06 model that we use in our work 
has stronger stellar-driven binary evolution than the K12 model. We also have smaller ratios $\frac{a_{\text{S06}}}{a_{\text{K12}}}$ 
when we hold the binary eccentricities fixed. This is because the binary eccentricities invariably grow when allowed to evolve, 
and lower eccentricities imply smaller GW-driven hardening rates. While an intuitive explanation of the difference between the S06 and K12 models 
is difficult to attain, the K12 work involves a more sophisticated, and possibly more realistic, treatment of the distribution of 
stellar orbits in the cores of merged galaxies than the S06 work. Given $f_{\text{orb}}\propto a^{-3/2}$ (Equation~(9)), 
a difference of a factor of two in the semi-major axes at which binary SMBH evolution begins to be GW-dominated 
corresponds to a difference of a factor of $2^{3/2}(\sim0.45\,{\rm dex})$ in orbital frequency.

\section{Effects on pulsar timing measurements of GW bursts from a binary SMBH}

Here, we provide a mathematical description of GW bursts from eccentric binary SMBHs. The spatial metric perturbation 
tensor, or GW strain, corresponding to a binary SMBH was defined by \citet{w87} to lowest order in the slow-motion, far-field regime using the 
quadrupole formula. This tensor, $h_{ij}$, can be written as \citep{w87}
\begin{equation}
h_{ij}=\sum_{S=+,\times}h^{S}e^{S}_{ij},
\end{equation}
where $e^{+}_{ij}$ and $e^{\times}_{ij}$ are the `plus' and `cross' polarisation tensors respectively, and $h^{+}$ and $h^{\times}$ are the 
time-varying polarisation amplitudes, which depend on the orbital phase $\theta$ (which is a function of time), the value of $\theta$ at the 
line of nodes ($\theta_{n}$), the value of $\theta$ at periastron ($\theta_{p}$), the orientation of the line of nodes ($\phi$), the binary inclination ($i$), 
and a factor 
\begin{equation}
A_{g}=\frac{4(GM_{C})^{5/3}}{c^{4}D(z)(1-e_{\text{GW}}^{2})}(2\pi f_{\text{orb}})^{2/3},
\end{equation}
where $D(z)$ is the comoving coordinate distance to redshift $z$ and $M_{C}=(M_{1}M_{2})^{3/5}(M_{1}+M_{2})^{-1/5}$ is the binary chirp mass.

GWs at the Earth and at a pulsar cause a fractional shift in the observed pulsar 
rotation frequency. Here, we neglect the effects of GWs at the pulsar, because, as outlined in, e.g., \citet{fl10}, GW bursts will generally affect pulsar 
timing data at vastly different times for different pulsars. This means that an approach that seeks to detect GW bursts by observing correlated effects in 
multiple pulsar datasets will only need to consider the effects of GWs at the Earth. For a pulsar with location defined by the unit direction tensor $p^{i}$, 
the observed fractional pulsar rotation frequency shift is given by \citep{w87,hjl+09}
\begin{equation}
\frac{\delta\nu_{p}}{\nu_{p}}=\frac{-p^{i}p^{j}h_{ij}}{2(1+\mu)},
\end{equation}
where $\mu$ is the cosine of the angle between the pulsar and GW source directions, and we follow the Einstein summation convention over the tensor indices.
 
Fractional shifts in $\nu_{p}$ will cause cumulative variations in ToAs. That is, 
\begin{equation}
R(t)=\int_{0}^{t}\frac{\delta\nu_{p}}{\nu_{p}}dt,
\end{equation}
where $R(t)$ is the ToA variation at a time $t$. In order to calculate the average duration and strength of a GW burst from a binary SMBH 
as manifested in $R(t)$, we need to calculate the variance 
\begin{equation}
\sigma_{R}^{2}(t)=\langle R^{2}(t) \rangle_{\theta_{n},\,\theta_{p},\,\phi,\,i,\,\alpha,\,\delta}
\end{equation}
where the angle brackets signify averaging over the subscripted quantities, and $\alpha$ and $\delta$ are the right ascension and declination of the 
pulsar assuming that the GW source is located along the z-axis. To simplify this, we set $\theta_{n}=\theta_{p}=0$, and express $R(t)$ as 
\begin{equation}
R(t)=\sum_{S=+,\times}R^{S}(t)G^{S},
\end{equation}
where 
\begin{equation}
R^{+,\times}(t)=\int_{0}^{t}h^{+,\times}dt
\end{equation}
and 
\begin{equation}
G^{+,\times}=-\frac{p^{i}p^{j}e^{+,\times}_{ij}}{2(1+\mu)}.
\end{equation}
The linear independence of the polarisation tensors implies that
\begin{equation}
\sigma_{R}^{2}(t)=\sum_{S=+,\times}\langle (R^{S}(t))^{2} \rangle_{\phi,\,i}\langle (G^{S})^{2} \rangle_{\alpha,\,\delta} .
\end{equation}
We find that 
\begin{equation}
\langle (G^{+,\times})^{2} \rangle_{\alpha,\,\delta}=\frac{1}{6}.
\end{equation}
Then, we have 
\begin{align}
\begin{split}
\sigma_{R}^{2}(t) = \,\,&\frac{A_{g}^{2}(1-e_{\text{GW}}^{2})^{3}\sin^{2}\theta}{720f_{\text{orb}}^{2}\pi^{3}(1+e_{\text{GW}}\cos\theta)^{2}} \\
&\times (3(8+e_{\text{GW}}^{2})+16e_{\text{GW}}\cos\theta+4\cos(2\theta)) \end{split} 
\end{align}
To summarise, Equation~(B11) gives the variance of the ToA variations caused by GWs from an individual eccentric binary SMBH at the Earth, 
averaged over all orientation parameters.

\bsp

\bibliographystyle{mn2e}
\bibliography{mn-jour,vikram}

\end{document}